\documentclass[aip,preprint]{revtex4-1}
\usepackage{graphicx}
\usepackage{amsmath}
\usepackage{physics}
\usepackage{ragged2e}
\begin{document}
\title{Proximity-induced Shiba states in a molecular junction}
\author{Joshua O. Island}
\email{jisland@physics.ucsb.edu}
\altaffiliation[Present address: ]{Department of Physics, University of California, Santa Barbara CA 93106 USA.}
\affiliation{Kavli Institute of Nanoscience, Delft University of Technology, Lorentzweg 1, 2628 CJ Delft, The Netherlands.}
\author{Rocco Gaudenzi}
\affiliation{Kavli Institute of Nanoscience, Delft University of Technology, Lorentzweg 1, 2628 CJ Delft, The Netherlands.}
\author{Joeri de Bruijckere}
\affiliation{Kavli Institute of Nanoscience, Delft University of Technology, Lorentzweg 1, 2628 CJ Delft, The Netherlands.}
\author{Enrique Burzur\'i}
\affiliation{Kavli Institute of Nanoscience, Delft University of Technology, Lorentzweg 1, 2628 CJ Delft, The Netherlands.}
\author{Carlos Franco}
\affiliation{Institut de Ci\'encia de Materials de Barcelona (ICMAB-CSIC) and CIBER-BBN, Campus de la UAB, 08193, Bellaterra, Spain.}
\author{Marta Mas-Torrent}
\affiliation{Institut de Ci\'encia de Materials de Barcelona (ICMAB-CSIC) and CIBER-BBN, Campus de la UAB, 08193, Bellaterra, Spain.}
\author{Concepci\'o Rovira}
\affiliation{Institut de Ci\'encia de Materials de Barcelona (ICMAB-CSIC) and CIBER-BBN, Campus de la UAB, 08193, Bellaterra, Spain.}
\author{Jaume Veciana}
\affiliation{Institut de Ci\'encia de Materials de Barcelona (ICMAB-CSIC) and CIBER-BBN, Campus de la UAB, 08193, Bellaterra, Spain.}
\author{Teun M. Klapwijk}
\affiliation{Kavli Institute of Nanoscience, Delft University of Technology, Lorentzweg 1, 2628 CJ Delft, The Netherlands.}
\affiliation{Physics Department, Moscow State Pedagogical University, Moscow 119991, Russia.}
\author{Ram\'on Aguado}
\affiliation{Instituto de Ciencia de Materiales de Madrid, Consejo Superior de Investigaciones Cient\'ificas (ICMM-CSIC), Sor Juana In\'es de la Cruz 3, 28049 Madrid, Spain.}
\author{Herre S. J. van der Zant}
\affiliation{Kavli Institute of Nanoscience, Delft University of Technology, Lorentzweg 1, 2628 CJ Delft, The Netherlands.}

\begin{abstract}
Superconductors containing magnetic impurities exhibit intriguing phenomena derived from the competition between Cooper pairing and Kondo screening. At the heart of this competition are the Yu-Shiba-Rusinov (Shiba) states which arise from the pair breaking effects a magnetic impurity has on a superconducting host. Hybrid superconductor-molecular junctions offer unique access to these states but the added complexity in fabricating such devices has kept their exploration to a minimum. Here, we report on the successful integration of a model spin 1/2 impurity, in the form of a neutral and stable all organic radical molecule, in proximity-induced superconducting break-junctions. Our measurements reveal excitations which are characteristic of a spin-induced Shiba state due to the radical's unpaired spin strongly coupled to a superconductor. By virtue of a variable molecule-electrode coupling, we access both the singlet and doublet ground states of the hybrid system which give rise to the doublet and singlet Shiba excited states, respectively. Our results show that Shiba states are a robust feature of the interaction between a paramagnetic impurity and a proximity-induced superconductor where the excited state is mediated by correlated electron-hole (Andreev) pairs instead of Cooper pairs.  
\end{abstract}
\maketitle

A quantum dot (QD) or impurity coupled to a superconductor constitutes a rich physical system in which many-body effects compete for the ground state\cite{de2010hybrid, franke2011competition, lee2012zero, chang2013tunneling, lee2014spin, lim2015shiba, hatter2015magnetic, vzonda2016perturbation, ruby2015tunneling}. The ground state can take the form of a BCS-like singlet, a spin degenerate doublet, or a Kondo-like singlet depending on the relative strengths of the characteristic energies of the competing phenomena (charging energy, $U$, superconducting gap, $\Delta$, Kondo energy, $k_BT_K$). For weak Coulomb interaction ($U << \Delta$), the BCS singlet, composed of the superposition of unoccupied and doubly occupied states of the dot, prevails. Subgap excitations of this ground state are the well-known Andreev bound states. This regime has been explored in carbon nanotube\cite{pillet2010andreev, pillet2013tunneling} and nanowire\cite{mourik2012signatures, chang2015hard} devices where $\Delta$ can be large enough relative to $U$ to allow the BCS superposition state.  For larger charging energy however, the doublet becomes the energetically favored ground state (at temperatures above $T_K$) and a competition between Kondo screening and Cooper pairing sets in at temperatures below $T_K$\cite{meng2009self, vzonda2016perturbation}. For weak Kondo energy ($k_BT_K<<\Delta$) the ground state is the degenerate doublet as screening is incomplete due to a lack of quasiparticles at the Fermi level. For strong Kondo energy ($k_BT_K>>\Delta$), quasiparticles screen the spin and the Kondo singlet becomes the ground state. Excitations on top of these ground states are the Yu-Shiba-Rusinov (or simply Shiba) states\cite{yu1965bound, shiba1968classical, rusinov1969superconductivity, balatsky2006impurity}. First experimentally observed in tunneling spectra of magnetic adatoms absorbed on Nb\cite{yazdani1997probing}, these states were recently shown to lead to topological Shiba bands required for the observation of Majorana end modes in atomic chains adsorbed on a superconducting surface\cite{nadj2013proposal, pientka2013topological, nadj2014observation}. 

Besides early tunneling experiments on Kondo alloys producing conventional Shiba bands\cite{dumoulin1977tunneling, machida1977proximity, machida1978proximity}, the great majority of investigations of superconductor-QD systems are with magnetic impurities\cite{yazdani1997probing, franke2011competition, hatter2015magnetic, menard2015coherent}, nanowires\cite{lee2012zero, chang2013tunneling} or nanotubes\cite{andersen2011nonequilibrium, kim2013transport, kumar2014temperature} coupled directly to a bulk superconductor. In these systems the singlet Shiba excited state is created by breaking a Cooper pair which allows a quasiparticle to pair with the localized spin in the quantized system. A similar interaction may occur through proximity induced superconductivity but has not been explored and requires attention as proposals and investigations of Majorana bound states engineered through proximity induced systems grow\cite{kitaev2001unpaired, alicea2010majorana, sau2010generic, oreg2010helical, alicea2012new, mourik2012signatures, albrecht2016exponential}. 

The hybrid superconductor-molecule device grants a unique exploration of the superconductor-QD phase diagram as the energy spacing between molecular orbitals of a molecule is typically orders of magnitude larger than the other energy scales completely suppressing the formation of the BCS-like singlet ground state and offering investigation of the large-$U$ regime. The added difficulty in fabricating such devices has limited their investigation to only a few studies\cite{winkelmann2009superconductivity, luo2009fabrication}. Winkelmann et al. have achieved contact to C$_{60}$ molecules using proximity induced gold and aluminum electrodes which showed the interplay of a spin-1/2 impurity coupled to intrinsically superconducting leads\cite{winkelmann2009superconductivity}. Gold is an attractive intermediary material because it allows coupling to a single molecule through a robust sulfur-gold bond leading to higher stability and stronger coupling.

Here we report on investigations of the large-$U$ regime of the superconductor-QD system and the observation of Shiba states in a completely proximitized superconducting junction hosting a model spin 1/2 impurity. As opposed to direct coupling to a bulk superconductor with a phase coherent condensate giving rise to spin induced Shiba states, Shiba states in our system are supported by correlated electron-hole (Andreev) pairs. Fig. 1(a) depicts this situation in which a magnetic impurity (in our case a molecule with an unpaired spin at its center signified by a red circle, see Supplemental Material for details\cite{SUPP1}\nocite{ballester1985inert}) on the left side is tunnel coupled to a proximity-induced superconductor on the right side. The proximity effect in the gold is mediated by the well-known Andreev reflection process that occurs at the interface between the normal metal (gold region) and the superconductor (blue region) giving rise to Andreev pairs in the gold leads\cite{andreev1964thermal} which can interact with the molecule through an exchange coupling. As a result of a variable molecule-electrode coupling for different devices, we access both the singlet and doublet ground states (and their corresponding Shiba excited states) of the large-$U$ superconductor-QD phase diagram. We further corroborate these results through calculations based on the Anderson impurity model while taking into account the proximity induced nature of the gold electrodes. 

\begin{figure}
\includegraphics [width=4.5in]{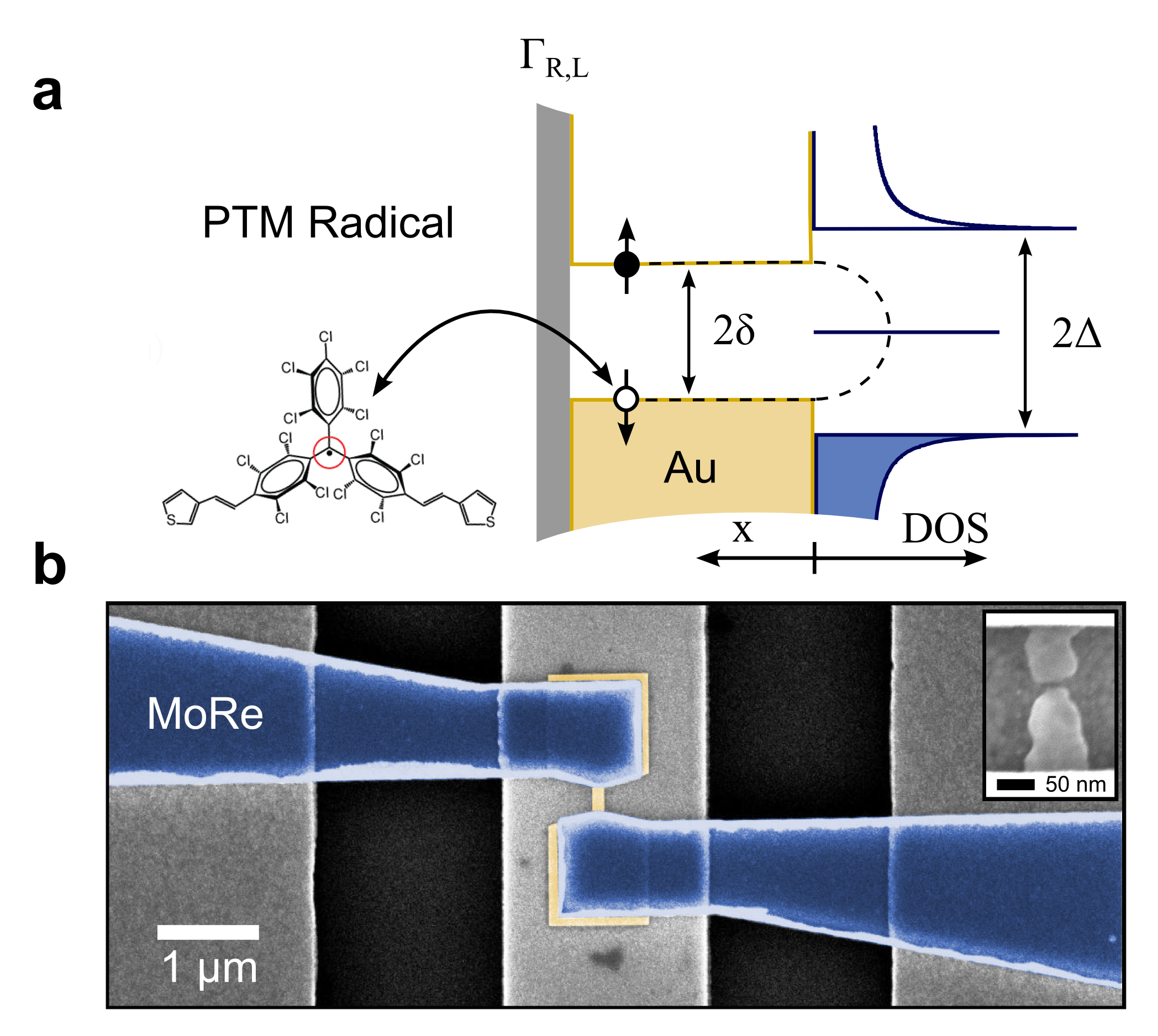}
\caption{\label{} \textbf{Proximity induced interaction with an all organic radical molecule and device design.} (a) Schematic representation of a radical molecule coupled to a proximity induced superconducting gold electrode. The red circle represents the unpaired spin at the center of the molecule. (b) False-colored scanning electron microscopy (SEM) image of a lithographically identical junction with superconducting electrodes colored blue and gold nanowire colored gold. The inset shows a SEM image of an electromigrated junction after measurement, scale bar 50 nm.}
\end{figure}

\section{Device design}
The junction fabrication and characterization have been detailed in Refs. \cite{gaudenzi2015superconducting, o2007self} which we briefly summarize here. Fig. 1(b) shows a false colored scanning electron microscopy (SEM) image of a representative junction. The blue colored electrodes are the bulk superconducting banks created by sputtering molybdenum-rhenium (60/40, $T_c \approx 8.5 K$). The gold colored nanowire and contact pads are created by electron-beam evaporation of gold (thickness, 12 nm). The gray vertical strip under the junction is a AlO$_x$/AuPd back gate which can be used to electrostatically gate the junction. A combination of electromigration and a self-breaking technique is used to open a nanogap in the gold nanowire. The inset shows an SEM image of a typical nanogap after electromigration. 

Several empty gaps are measured at low temperatures in order to characterize the superconducting proximity effect in the gold electrodes without molecules (see Supplemental Material for details\cite{SUPP2}\nocite{wolz2011evidence, belzig1996local}). In order to carry out measurements on the polychlorotriphenylmethyl (PTM) molecule (see Supplemental Material for details\cite{SUPP1}), a solution of molecules (concentration 1 mM) is drop-cast onto an array of 24 junctions before electromigration. After the electromigration and self-breaking of all junctions, the solution is evaporated leaving behind roughly a monolayer of molecules and the sample is cooled down in a dilution refrigerator (base temperature 100 mK). 

We find that the normal state characteristics (and Kondo effect) of the molecular junctions are consistent with previous reports of the same molecule in completely gold break junctions\cite{frisenda2015kondo} (see Supplemental Materials for details\cite{SUPP3}\nocite{parks2007tuning, costi1994transport}). Furthermore, from the linear conductance in the Kondo regime we estimate an asymmetry in the couplings ($\Gamma_L, \Gamma_R$) to the leads of $\Gamma_L/\Gamma_R\approx 5 \times 10^{-4}$.

\section{The superconducting state}
Moving now to the characteristics in the superconducting state, the black curve in fig. 2(a) shows the differential conductance as a function of voltage bias in the superconducting state ($B = 0$ T) for device D. The red curve in the same panel is the measurement in the normal state ($B = 200$ mT) showing the zero bias peak arising from the Kondo effect discussed in the Supplemental Materials\cite{SUPP3}. In the superconducting state two peaks are discernible, symmetric in bias, accompanied by side dips at higher bias which are qualitatively distinctive from the empty junction curves shown in the Supplemental Material\cite{SUPP2}. These peaks signal excitations of the coupled superconductor-QD system and have been shown to be associated with (multiple) Andreev reflections (MAR) and spin induced Yu-Shiba-Rusniov states. Conductance peaks due to MAR however are not accompanied by side dips and become less probable as the asymmetry between the left and right leads is increased\cite{andersen2011nonequilibrium, lee2012zero}. Asymmetries of $\Gamma_L/\Gamma_R\approx 10^{-2}-10^{-3}$ in nanowire experiments are enough to completely suppress MAR contributions\cite{lee2012zero}. With the built in high asymmetry of the couplings and given the spin 1/2 Kondo effect in the normal state, a natural explanation for the excitation peaks in our hybrid molecular devices is that they originate from spin induced Shiba states. 

Figure 2(c) shows a schematic of the coupling situation and density of states. The blue colored electrode and corresponding DOS represent the hybridization of the radical with the more strongly coupled electrode resulting in Shiba excited states. Microscopically, these states arise from the interaction of Andreev pairs at finite energy in the proximity-induced lead and the unpaired spin of the radical molecule. The weakly coupled gold lead on the left side probes the hybridization of the molecule with the right electrode. This results in conductance peaks in fig. 2(a) at energies of $\pm(E_b\pm\delta_L)$ where $E_b$ is the excited state energy and $\delta_L$ is the proximity-induced gap of the probe electrode. In addition to these peaks, side shoulders are visible at lower bias voltage which we interpret as Shiba replicas. Shiba replicas are visible for a sufficient density of quasiparticles in the mini-gap of the probe electrode which is reasonable considering the soft proximity-induced gap of our empty junctions (see Supplemental Materials\cite{SUPP2}). With an increase in temperature we furthermore observe the emergence of an anomalous zero-bias peak (dashed curve in Fig. 2(a)) which we interpret as a mini-Kondo due to increased quasiparticle filling (see Supplemental Material\cite{SUPP4}). 

A comparison of the characteristic energies ($\delta_{avg}$ vs. $k_BT_K$) for device D allows us to determine the ground state of the coupled system. The edge of the dips in fig. 2(a) roughly correspond to $\delta_L + \delta_R$ from which we calculate an average induced gap of $\delta_{avg} = (\delta_L+\delta_R)/2 = 0.8$ meV. Compared with the Kondo energy extracted from the normal state temperature dependence of the Kondo peak ($k_BT_K =  0.2$ meV) (see Supplemental Materials\cite{SUPP3}), we find that the average induced gap is 4 times larger. In this device, the larger finite Andreev pair energy results in a doublet ground state where the radical's spin is not sufficiently screened by quasiparticle states near the Fermi energy in the superconducting state to allow the Kondo singlet. Application of a perpendicular magnetic field however, allows us to tune the ground state from the doublet to the Kondo singlet (see inset of fig. 2(a)).  

Our low temperature measurements are further supported by calculations. Starting with a modified Anderson impurity model (Anderson Hamiltonian coupled to leads modeled by BCS Hamiltonians) we calculate the differential conductance as a function of bias voltage using the non-crossing approximation\cite{lee2012zero, clerk2000loss, sellier2005pi} (see Supplemental Materials for details\cite{SUPP5}\nocite{langreth1991derivation, wingreen1994anderson, hettler1994nonlinear, hettler1998nonequilibrium, aguado2003kondo, costi1996spectral, devreeese2013linear, meir1992landauer}
). The broadening of the DOS due to the proximity effect is handled by introducing the Dynes function which is taken as a phenomenological broadening term that softens the BCS DOS of the bulk reservoirs\cite{lee2012zero, dynes1978direct}. Fig. 2(b) shows a conductance curve from this calculation where the main features of the experimental curve can be reproduced. A zero-bias peak is present when the order parameter in the leads is set to zero (red curve). In the superconducting state with $\delta=4k_BT_K$ (for calculations $\delta_L = \delta_R = \delta$) and broadening (0.15$\delta$), two peaks can be seen with accompanying side dips at higher bias (black curve). Note that we have also incorporated an asymmetry in the lead couplings of $\Gamma_L/\Gamma_R=1/2$ to better model the results. Quantitative differences are primarily due to the exact shape of the DOS for each gold lead. Differences in the lengths of the two gold leads (see Supplemental Material for details\cite{SUPP2}) mean the broadening for the two leads would be different ($\delta_L \neq \delta_R$). Additionally, the shape of the gold lead near the molecular junction ultimately plays a role in the proximity effect and thus the shape of the DOS which would require more intensive calculations. 

\begin{figure}
\includegraphics [width=\linewidth]{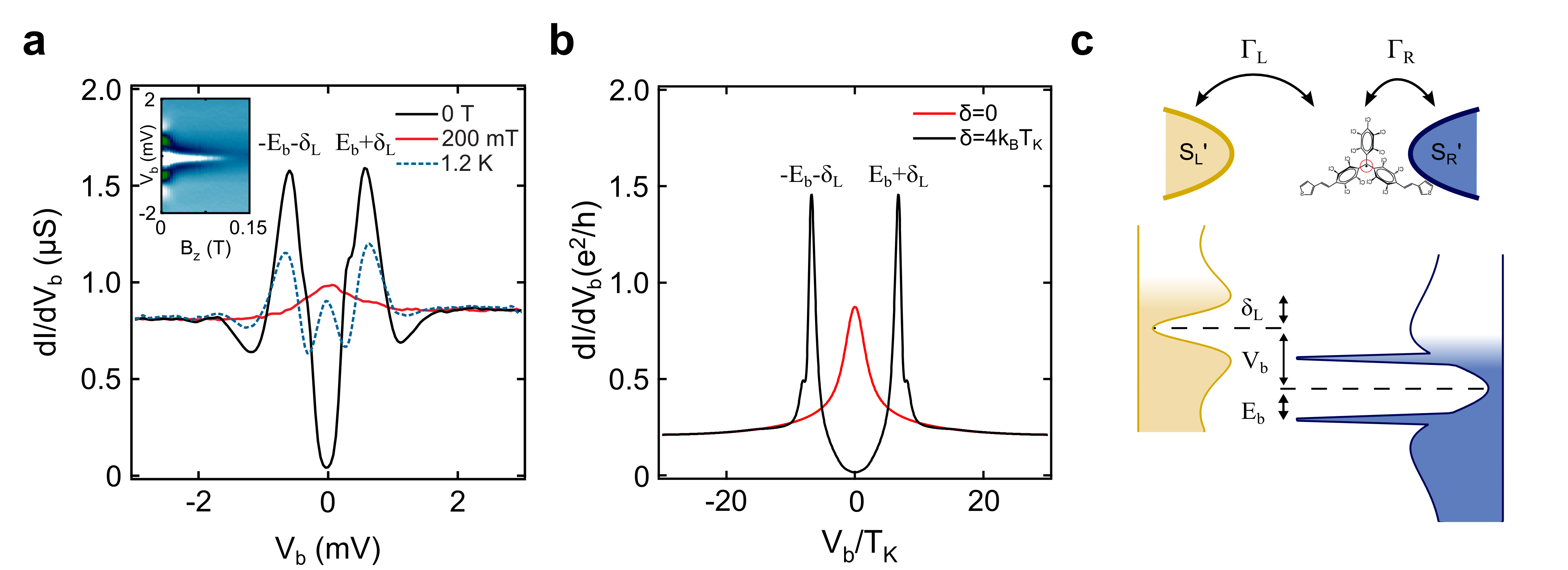}
\caption{\label{}  \textbf{Characterization and calculations of a molecular junction in the superconducting state.} (a) Low temperature (100 mK) measurement of the differential conductance ($dI/dV_b$) as a function of voltage bias ($V_b$) of a radical molecular junction at zero magnetic field (superconducting state) and 200 mT (normal state). The blue dashed curve shows the same measurement at 1.2 K ($B = 0$ T). The inset shows the magnetic field dependence of the Shiba peaks. (b) Theoretical calculations of the modeled system taking into account the proximity-induced DOS of the leads. (c) Schematic representation of the probing of the Shiba excited state as a result of the coupling of the radical molecule with the proximity-induced superconducting Au electrode.}
\end{figure}

\section{Ground state spectroscopy in the superconducting state}
Depending on the relative energies of the proximity-induced gap and the Kondo energy, the ground state of the system, even in the superconducting state, can take the form of a doublet for weak coupling ($k_BT_K < \delta_{avg}$) or a Kondo-like singlet for strong coupling ($k_BT_K > \delta_{avg}$). By virtue of a variable molecule-electrode coupling, we probe a large range of Kondo energies relative to the average induced gap energy. In the Supplemental Material\cite{SUPP6} we show an overview of the 7 devices (A-G) having similar characteristics ordered by increasing Kondo energy. The Kondo temperature ranges from ($\approx 1$ K to $18$ K) for the seven devices. 
\begin{figure}
\includegraphics [width=\linewidth]{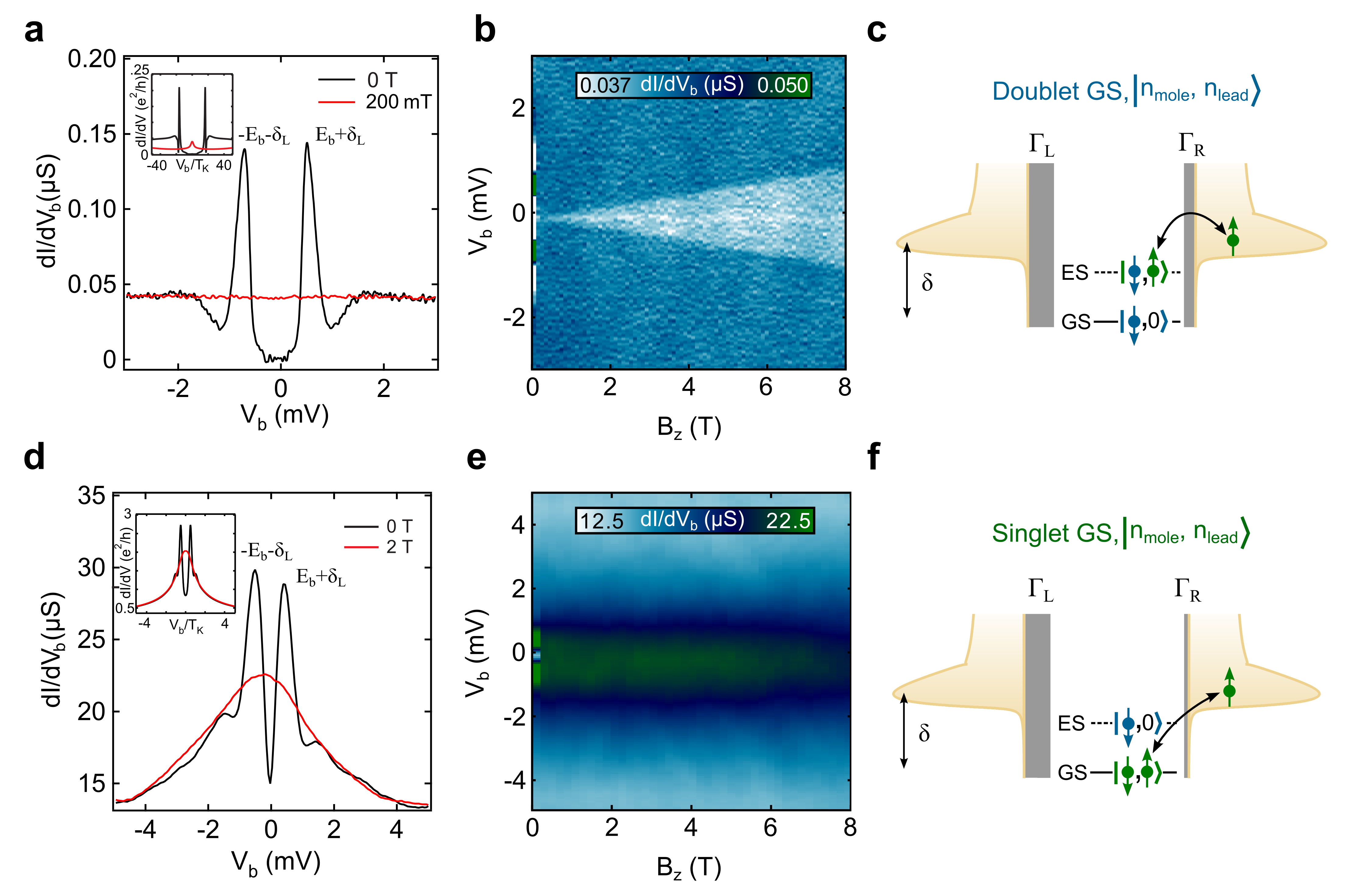}
\caption{\label{} \textbf{Ground state spectroscopy of the hybrid radical system.} (a) The differential conductance ($dI/dV_b$) as a function of voltage bias ($V_b$) of a radical molecular junction at zero magnetic field (superconducting state) and 200 mT (normal state) in the doublet ground state regime ($k_BT_K<\delta_{avg}$). Inset shows calculations for this regime ($\delta=9k_BT_K$). (b) Color plot of the differential conductance ($dI/dV_b$) as a function of voltage bias and magnetic field for the junction in panel (a). (c) Excitation picture of the ground and excited states in the doublet ground state regime. (d) The differential conductance ($dI/dV_b$) as a function of voltage bias ($V_b$) of a radical molecular junction at zero magnetic field (superconducting state) and 2 T (normal state) in the singlet ground state regime ($k_BT_K>\delta_{avg}$). Inset shows calculations for this regime ($\delta=0.5k_BT_K$). (e) Color plot of the differential conductance ($dI/dV_b$) as a function of voltage bias and magnetic field for the junction in panel (d). (f) Excitation picture of the ground and excited states in the singlet ground state regime.}
\end{figure}

In fig. 3 we show the weakest and strongest Kondo energy devices (junctions A and G). For junction A a Kondo peak cannot be discerned in the normal state (red curve fig. 3(a)) but a splitting ($g = 2.1$) of the background is observed at higher magnetic fields fig. 3(b)). We estimate a Kondo temperature of $\approx 1$ K from the critical field ($B_c \approx 0.5k_BT_K/(g\mu_B)$) at which the background begins to split in fig. 3(b) ($B_c \approx 0.36$ T). Following the analysis for device D above we estimate an average induced gap of $0.8$ meV. Comparing the two energy scales (0.8 meV vs. 0.09 meV), this device is similar to device D in which the doublet ground state wins over the Kondo singlet. The inset of fig. 3(a) shows the calculations for this regime where we have taken $\delta=9k_BT_K$ and built in an asymmetry of $\Gamma_R/\Gamma_L=20$ to better simulate the results. Shiba peaks in this regime correspond to singlet excitations of the doublet ground state. This situation is depicted in fig. 3(c) in the excitation picture where a thick barrier and reduced quasiparticle states near the Fermi energy prevent the Kondo singlet from claiming the ground state of the coupled system. Here we signify the ground state and the excited state by occupancies of the molecule and the leads, $\ket{n_{mole}, n_{lead}}$. The Shiba excited state for this ground state, $\ket{\downarrow, \uparrow}$, is a singlet composed of the molecule's unpaired spin and an electron from a correlated Andreev pair in the leads (signified by the black arrow). 

In fig. 3(d) we present the device with the largest Kondo temperature observed (18 K, estimated from the peak width). Comparing the characteristic energies for this device ($\delta_{avg} = 0.7$ meV, $k_BT_K = 1.6$ meV) we find that the large Kondo energy allows for screening of the radical's spin even in the superconducting state. This results in a singlet ground state at $B = 0$ T. Calculations for this regime are shown in the inset of fig. 3(d) where we have taken $\delta=0.5k_BT_K$ and a broadening of 0.25$\delta$. Excitations of this ground state are the magnetic doublet. This is depicted in fig. 3(f) in the excitation picture. A weak barrier allows for screening from electrons in the proximity-induced leads and a singlet state is formed by the combination of the radical's spin and an electron in the leads. 

Finally, using the average proximity induced gap, we make a qualitative estimate of the bound state energies for the 7 devices and extract the approximate ratios of $k_BT_K/\delta_{avg}$ for which a quantum phase transition occurs from the doublet ground state to the singlet ground state (see the Supplemental Materials for details\cite{SUPP7}). Of the 7 devices studied, four were found to have induced superconducting gaps with energies greater than their respective Kondo energy and three were found to have energies less than their Kondo energy. Taken all together, the data suggests a quantum phase transition occurs for $k_BT_K/\delta_{avg}$ between 0.9 and 1.1. This is in agreement with STM studies of molecules on a Pb surface ($k_BT_K/\Delta\approx1$)\cite{franke2011competition} but slightly higher than early NRG theory calculations which predict a transition at $k_BT_K/\Delta\approx0.3$ (Ref. \cite{sakai1993numerical}) and more recent investigations in nanowire systems showing quantitative agreement between experiment and NRG calculations for a phase transition occurring at $k_BT_K/\Delta\approx0.6$ (Ref. \cite{lee2016scaling}). 

\section{Conclusion}
In conclusion, we have presented an investigation of the large-$U$ superconductor-QD phase diagram in the form of a radical molecule coupled to a proximity induced superconductor. In the superconducting state, we observe excitations which are characteristic of Shiba states as a result of the coupling between the radical and a proximity-induced superconductor. By applying a finite magnetic field, the proximity effect can be suppressed which allows a spin 1/2 Kondo effect. For the devices with the weakest and strongest Kondo energies we are able to probe both the doublet and singlet ground states which give rise to the singlet and doublet Shiba excited states, respectively. Finally, we estimate that the quantum phase transition from the doublet to the singlet ground state occurs for ratios of $k_BT_K$ to $\delta_{avg}$ from 0.9 to 1.1. These findings are supported by calculations of the Anderson impurity model modified to account for the proximity-induced superconducting electrodes. Hybrid molecular junctions offer a unique investigation of the superconductor-QD system. In particular, with a suitable choice of molecule, an \textit{in-situ} tunable Kondo effect would allow direct driving of the singlet to doublet quantum phase transition for a spin 1/2 impurity and extraction of the energies at which it occurs\cite{franke2011competition, matsuura1977effects, sakai1993numerical}. 

\section{Acknowledgments}
The authors thank Jens Paaske for insightful discussions. The authors acknowledge financial support by the Dutch Organization for Fundamental research (FOM), the Ministry of Education, Culture, and Science (OCW), the Netherlands Organization for Scientific Research (NWO), and an ERC advanced grant (MolS@MolS). C.F., M.M.-T., C.R. and J.V. acknowledge the financial support from Spanish Ministry of Economy and Competitiveness, through the DGI (CTQ2013-40480-R) and ‘Severo Ochoa’ Programme for Centres of Excellence in R\&D (SEV-2015-0496), the Generalitat de Catalunya (2014SGR-17) and the Networking Research Center of Bioengineering, Biomaterials and Nanomedicine (CIBER-BBN). C.F. is enrolled in the Materials Science PhD program of UAB. T.M.K. acknowledges the support from the European Research Council Advanced Grant No. 339306 (METIQUM) and from the Ministry of Education and Science of the Russian Federation, Contract No. 14.B25.31.0007. R.A. acknowledges grants FIS2012-33521 and FIS-2015-64654 (MINECO/FEDER).

\clearpage

\bibliography{Island_Shiba_ref}

\begin{thebibliography}{10}

\bibitem{de2010hybrid}
Silvano De~Franceschi, Leo Kouwenhoven, Christian Sch{\"o}nenberger, and
  Wolfgang Wernsdorfer.
\newblock Hybrid superconductor-quantum dot devices.
\newblock {\em Nature Nanotechnology}, 5(10):703--711, 2010.

\bibitem{franke2011competition}
KJ~Franke, G~Schulze, and JI~Pascual.
\newblock Competition of superconducting phenomena and kondo screening at the
  nanoscale.
\newblock {\em Science}, 332(6032):940--944, 2011.

\bibitem{lee2012zero}
Eduardo~JH Lee, Xiaocheng Jiang, Ram{\'o}n Aguado, Georgios Katsaros, Charles~M
  Lieber, and Silvano De~Franceschi.
\newblock Zero-bias anomaly in a nanowire quantum dot coupled to
  superconductors.
\newblock {\em Physical Review Letters}, 109(18):186802, 2012.

\bibitem{chang2013tunneling}
W~Chang, VE~Manucharyan, Thomas~Sand Jespersen, Jesper Nyg{\aa}rd, and
  Charles~M Marcus.
\newblock Tunneling spectroscopy of quasiparticle bound states in a spinful
  josephson junction.
\newblock {\em Physical Review Letters}, 110(21):217005, 2013.

\bibitem{lee2014spin}
Eduardo~JH Lee, Xiaocheng Jiang, Manuel Houzet, Ram{\'o}n Aguado, Charles~M
  Lieber, and Silvano De~Franceschi.
\newblock Spin-resolved andreev levels and parity crossings in hybrid
  superconductor-semiconductor nanostructures.
\newblock {\em Nature nanotechnology}, 9(1):79--84, 2014.

\bibitem{lim2015shiba}
Jong~Soo Lim, Rosa L{\'o}pez, Ram{\'o}n Aguado, et~al.
\newblock Shiba states and zero-bias anomalies in the hybrid
  normal-superconductor anderson model.
\newblock {\em Physical Review B}, 91(4):045441, 2015.

\bibitem{hatter2015magnetic}
Nino Hatter, Benjamin~W Heinrich, Michael Ruby, Jose~I Pascual, and Katharina~J
  Franke.
\newblock Magnetic anisotropy in shiba bound states across a quantum phase
  transition.
\newblock {\em Nature communications}, 6:8988, 2015.

\bibitem{vzonda2016perturbation}
Martin {\v{Z}}onda, Vladislav Pokorn{\`y}, V{\'a}clav Jani{\v{s}}, and
  Tom{\'a}{\v{s}} Novotn{\`y}.
\newblock Perturbation theory for an anderson quantum dot asymmetrically
  attached to two superconducting leads.
\newblock {\em Physical Review B}, 93(2):024523, 2016.

\bibitem{ruby2015tunneling}
Michael Ruby, Falko Pientka, Yang Peng, Felix von Oppen, Benjamin~W Heinrich,
  and Katharina~J Franke.
\newblock Tunneling processes into localized subgap states in superconductors.
\newblock {\em Physical Review Letters}, 115(8):087001, 2015.

\bibitem{pillet2010andreev}
JD~Pillet, CHL Quay, P~Morfin, C~Bena, A~Levy Yeyati, and P~Joyez.
\newblock Andreev bound states in supercurrent-carrying carbon nanotubes
  revealed.
\newblock {\em Nature Physics}, 6(12):965--969, 2010.

\bibitem{pillet2013tunneling}
J-D Pillet, P~Joyez, MF~Goffman, et~al.
\newblock Tunneling spectroscopy of a single quantum dot coupled to a
  superconductor: From kondo ridge to andreev bound states.
\newblock {\em Physical Review B}, 88(4):045101, 2013.

\bibitem{mourik2012signatures}
Vincent Mourik, Kun Zuo, Sergey~M Frolov, SR~Plissard, EPAM Bakkers, and
  LP~Kouwenhoven.
\newblock Signatures of majorana fermions in hybrid
  superconductor-semiconductor nanowire devices.
\newblock {\em Science}, 336(6084):1003--1007, 2012.

\bibitem{chang2015hard}
W~Chang, SM~Albrecht, TS~Jespersen, Ferdinand Kuemmeth, P~Krogstrup,
  J~Nyg{\aa}rd, and CM~Marcus.
\newblock Hard gap in epitaxial semiconductor--superconductor nanowires.
\newblock {\em Nature nanotechnology}, 10(3):232--236, 2015.

\bibitem{meng2009self}
Tobias Meng, Serge Florens, and Pascal Simon.
\newblock Self-consistent description of andreev bound states in josephson
  quantum dot devices.
\newblock {\em Physical Review B}, 79(22):224521, 2009.

\bibitem{yu1965bound}
L~Yu.
\newblock Bound state in superconductors with paramagnetic impurities.
\newblock {\em Acta Phys. Sin}, 21:75--91, 1965.

\bibitem{shiba1968classical}
Hiroyuki Shiba.
\newblock Classical spins in superconductors.
\newblock {\em Progress of theoretical Physics}, 40(3):435--451, 1968.

\bibitem{rusinov1969superconductivity}
AI~Rusinov.
\newblock Superconductivity near a paramagnetic impurity.
\newblock {\em Soviet Journal of Experimental and Theoretical Physics Letters},
  9:85, 1969.

\bibitem{balatsky2006impurity}
AV~Balatsky, I~Vekhter, and Jian-Xin Zhu.
\newblock Impurity-induced states in conventional and unconventional
  superconductors.
\newblock {\em Reviews of Modern Physics}, 78(2):373, 2006.

\bibitem{yazdani1997probing}
Ali Yazdani, BA~Jones, CP~Lutz, MF~Crommie, and DM~Eigler.
\newblock Probing the local effects of magnetic impurities on
  superconductivity.
\newblock {\em Science}, 275(5307):1767--1770, 1997.

\bibitem{nadj2013proposal}
S~Nadj-Perge, IK~Drozdov, BA~Bernevig, and Ali Yazdani.
\newblock Proposal for realizing majorana fermions in chains of magnetic atoms
  on a superconductor.
\newblock {\em Physical Review B}, 88(2):020407, 2013.

\bibitem{pientka2013topological}
Falko Pientka, Leonid~I Glazman, and Felix von Oppen.
\newblock Topological superconducting phase in helical shiba chains.
\newblock {\em Physical Review B}, 88(15):155420, 2013.

\bibitem{nadj2014observation}
Stevan Nadj-Perge, Ilya~K Drozdov, Jian Li, Hua Chen, Sangjun Jeon, Jungpil
  Seo, Allan~H MacDonald, B~Andrei Bernevig, and Ali Yazdani.
\newblock Observation of majorana fermions in ferromagnetic atomic chains on a
  superconductor.
\newblock {\em Science}, 346(6209):602--607, 2014.

\bibitem{dumoulin1977tunneling}
L~Dumoulin, E~Guyon, and P~Nedellec.
\newblock Tunneling study of localized bands in superconductors with magnetic
  impurities (normal kondo alloys in the superconducting proximity).
\newblock {\em Physical Review B}, 16(3):1086, 1977.

\bibitem{machida1977proximity}
Kazushige Machida.
\newblock Proximity effect for superconductors containing transition metal
  impurities. i.
\newblock {\em Journal of Low Temperature Physics}, 27(5-6):737--745, 1977.

\bibitem{machida1978proximity}
Kazushige Machida and L~Dumoulin.
\newblock Proximity effect for superconductors containing transition metal
  impurities. ii.
\newblock {\em Journal of Low Temperature Physics}, 31(1-2):143--152, 1978.

\bibitem{menard2015coherent}
Gerbold~C M{\'e}nard, S{\'e}bastien Guissart, Christophe Brun, St{\'e}phane
  Pons, Vasily~S Stolyarov, Fran{\c{c}}ois Debontridder, Matthieu~V Leclerc,
  Etienne Janod, Laurent Cario, Dimitri Roditchev, et~al.
\newblock Coherent long-range magnetic bound states in a superconductor.
\newblock {\em Nature Physics}, 11(12):1013--1016, 2015.

\bibitem{andersen2011nonequilibrium}
Brian~M{\o}ller Andersen, Karsten Flensberg, Verena Koerting, and Jens Paaske.
\newblock Nonequilibrium transport through a spinful quantum dot with
  superconducting leads.
\newblock {\em Physical Review Letters}, 107(25):256802, 2011.

\bibitem{kim2013transport}
Bum-Kyu Kim, Ye-Hwan Ahn, Ju-Jin Kim, Mahn-Soo Choi, Myung-Ho Bae, Kicheon
  Kang, Jong~Soo Lim, Rosa L{\'o}pez, and Nam Kim.
\newblock Transport measurement of andreev bound states in a kondo-correlated
  quantum dot.
\newblock {\em Physical Review Letters}, 110(7):076803, 2013.

\bibitem{kumar2014temperature}
A~Kumar, M~Gaim, D~Steininger, A~Levy Yeyati, A~Mart{\'\i}n-Rodero,
  AK~H{\"u}ttel, and C~Strunk.
\newblock Temperature dependence of andreev spectra in a superconducting carbon
  nanotube quantum dot.
\newblock {\em Physical Review B}, 89(7):075428, 2014.

\bibitem{kitaev2001unpaired}
A~Yu Kitaev.
\newblock Unpaired majorana fermions in quantum wires.
\newblock {\em Physics-Uspekhi}, 44(10S):131, 2001.

\bibitem{alicea2010majorana}
Jason Alicea.
\newblock Majorana fermions in a tunable semiconductor device.
\newblock {\em Physical Review B}, 81(12):125318, 2010.

\bibitem{sau2010generic}
Jay~D Sau, Roman~M Lutchyn, Sumanta Tewari, and S~Das Sarma.
\newblock Generic new platform for topological quantum computation using
  semiconductor heterostructures.
\newblock {\em Physical review letters}, 104(4):040502, 2010.

\bibitem{oreg2010helical}
Yuval Oreg, Gil Refael, and Felix von Oppen.
\newblock Helical liquids and majorana bound states in quantum wires.
\newblock {\em Physical review letters}, 105(17):177002, 2010.

\bibitem{alicea2012new}
Jason Alicea.
\newblock New directions in the pursuit of majorana fermions in solid state
  systems.
\newblock {\em Reports on Progress in Physics}, 75(7):076501, 2012.

\bibitem{albrecht2016exponential}
SM~Albrecht, AP~Higginbotham, M~Madsen, F~Kuemmeth, TS~Jespersen, Jesper
  Nyg{\aa}rd, P~Krogstrup, and CM~Marcus.
\newblock Exponential protection of zero modes in majorana islands.
\newblock {\em Nature}, 531(7593):206--209, 2016.

\bibitem{winkelmann2009superconductivity}
Clemens~B Winkelmann, Nicolas Roch, Wolfgang Wernsdorfer, Vincent Bouchiat, and
  Franck Balestro.
\newblock Superconductivity in a single-c60 transistor.
\newblock {\em Nature Physics}, 5(12):876--879, 2009.

\bibitem{luo2009fabrication}
Kang Luo and Zhen Yao.
\newblock Fabrication of nanometer-spaced superconducting pb electrodes.
\newblock {\em Applied Physics Letters}, 95(11):113115, 2009.

\bibitem{SUPP1}
See Supplemental Material for a complete description of the molecule which
  includes Ref. \cite{ballester1985inert}.

\bibitem{ballester1985inert}
Manuel Ballester.
\newblock Inert free radicals (ifr): a unique trivalent carbon species.
\newblock {\em Accounts of Chemical Research}, 18(12):380--387, 1985.

\bibitem{andreev1964thermal}
AF~Andreev.
\newblock The thermal conductivity of the intermediate state in
  superconductors.
\newblock {\em Soviet Physics JETP}, 19(5):1823, 1964.

\bibitem{gaudenzi2015superconducting}
R~Gaudenzi, JO~Island, J~De~Bruijckere, E~Burzur{\'\i}, TM~Klapwijk, and HSJ
  Van~der Zant.
\newblock Superconducting molybdenum-rhenium electrodes for single-molecule
  transport studies.
\newblock {\em Applied Physics Letters}, 106(22):222602, 2015.

\bibitem{o2007self}
K~O’Neill, EA~Osorio, and HSJ Van~der Zant.
\newblock Self-breaking in planar few-atom au constrictions for
  nanometer-spaced electrodes.
\newblock {\em Applied Physics Letters}, 90(13):133109, 2007.

\bibitem{SUPP2}
See Supplemental Material for low temperature characteristics of an empty gap
  which includes Refs. \cite{wolz2011evidence, belzig1996local}.

\bibitem{wolz2011evidence}
Michael Wolz, Christian Debuschewitz, Wolfgang Belzig, and Elke Scheer.
\newblock Evidence for attractive pair interaction in diffusive gold films
  deduced from studies of the superconducting proximity effect with aluminum.
\newblock {\em Physical Review B}, 84(10):104516, 2011.

\bibitem{belzig1996local}
Wolfgang Belzig, Christoph Bruder, and Gerd Sch{\"o}n.
\newblock Local density of states in a dirty normal metal connected to a
  superconductor.
\newblock {\em Physical Review B}, 54(13):9443, 1996.

\bibitem{frisenda2015kondo}
Riccardo Frisenda, Rocco Gaudenzi, Carlos Franco, Marta Mas-Torrent,
  Concepci{\'o} Rovira, Jaume Veciana, Isaac Alcon, Stefan~T Bromley, Enrique
  Burzur{\'\i}, and Herre~SJ Van~der Zant.
\newblock Kondo effect in a neutral and stable all organic radical single
  molecule break junction.
\newblock {\em Nano letters}, 15(5):3109--3114, 2015.

\bibitem{SUPP3}
See Supplemental Material for normal state characterization of a molecular
  junction which includes Refs. \cite{parks2007tuning, costi1994transport}.

\bibitem{parks2007tuning}
JJ~Parks, AR~Champagne, GR~Hutchison, S~Flores-Torres, HD~Abruna, and DC~Ralph.
\newblock Tuning the kondo effect with a mechanically controllable break
  junction.
\newblock {\em Physical Review Letters}, 99(2):026601, 2007.

\bibitem{costi1994transport}
TA~Costi, AC~Hewson, and V~Zlatic.
\newblock Transport coefficients of the anderson model via the numerical
  renormalization group.
\newblock {\em Journal of Physics: Condensed Matter}, 6(13):2519, 1994.

\bibitem{SUPP4}
See Supplemental Material for temperature dependence of the mini-Kondo along
  with model calculations.

\bibitem{clerk2000loss}
Aashish~A Clerk and Vinay Ambegaokar.
\newblock Loss of $\pi$-junction behavior in an interacting impurity josephson
  junction.
\newblock {\em Physical Review B}, 61(13):9109, 2000.

\bibitem{sellier2005pi}
Gabriel Sellier, Thilo Kopp, Johann Kroha, and Yuri~S Barash.
\newblock $\pi$ junction behavior and andreev bound states in kondo quantum
  dots with superconducting leads.
\newblock {\em Physical Review B}, 72(17):174502, 2005.

\bibitem{SUPP5}
See Supplemental Material for a complete description of the theoretical model
  which includes Refs. \cite{langreth1991derivation, wingreen1994anderson,
  hettler1994nonlinear, hettler1998nonequilibrium, aguado2003kondo,
  costi1996spectral, devreeese2013linear, meir1992landauer}.

\bibitem{langreth1991derivation}
David~C Langreth and Peter Nordlander.
\newblock Derivation of a master equation for charge-transfer processes in
  atom-surface collisions.
\newblock {\em Physical Review B}, 43(4):2541, 1991.

\bibitem{wingreen1994anderson}
Ned~S Wingreen and Yigal Meir.
\newblock Anderson model out of equilibrium: Noncrossing-approximation approach
  to transport through a quantum dot.
\newblock {\em Physical review B}, 49(16):11040, 1994.

\bibitem{hettler1994nonlinear}
Matthias~H Hettler, Johann Kroha, and Selman Hershfield.
\newblock Nonlinear conductance for the two channel anderson model.
\newblock {\em Physical review letters}, 73(14):1967, 1994.

\bibitem{hettler1998nonequilibrium}
Matthias~H Hettler, Johann Kroha, and Selman Hershfield.
\newblock Nonequilibrium dynamics of the anderson impurity model.
\newblock {\em Physical Review B}, 58(9):5649, 1998.

\bibitem{aguado2003kondo}
Ram{\'o}n Aguado and David~C Langreth.
\newblock Kondo effect in coupled quantum dots: A noncrossing approximation
  study.
\newblock {\em Physical Review B}, 67(24):245307, 2003.

\bibitem{costi1996spectral}
TA~Costi, J~Kroha, and P~W{\"o}lfle.
\newblock Spectral properties of the anderson impurity model: Comparison of
  numerical-renormalization-group and noncrossing-approximation results.
\newblock {\em Physical Review B}, 53(4):1850, 1996.

\bibitem{devreeese2013linear}
J~Devreeese.
\newblock {\em Linear and nonlinear electron transport in solids}, volume~17.
\newblock Springer Science \& Business Media, 1976.

\bibitem{meir1992landauer}
Yigal Meir and Ned~S Wingreen.
\newblock Landauer formula for the current through an interacting electron
  region.
\newblock {\em Physical review letters}, 68(16):2512, 1992.

\bibitem{dynes1978direct}
RC~Dynes, V~Narayanamurti, and J~Pm Garno.
\newblock Direct measurement of quasiparticle-lifetime broadening in a
  strong-coupled superconductor.
\newblock {\em Physical Review Letters}, 41(21):1509, 1978.

\bibitem{SUPP6}
See Supplemental Material for an overview of the seven molecular devices having
  similar characteristics.

\bibitem{SUPP7}
See Supplemental Material for a plot of the Shiba energy ($E_b$) versus the
  Kondo energy ($k_BT_K$) for all seven molecular devices.

\bibitem{sakai1993numerical}
Osamu Sakai, Yukihiro Shimizu, Hiroyuki Shiba, and Koji Satori.
\newblock Numerical renormalization group study of magnetic impurities in
  superconductors. ii. dynamical excitation spectra and spatial variation of
  the order parameter.
\newblock {\em Journal of the Physical Society of Japan}, 62(9):3181--3197,
  1993.

\bibitem{lee2016scaling}
Eduardo~JH Lee, Xiaocheng Jiang, Rok Zitko, Ramon Aguado, Charles~M Lieber, and
  Silvano De~Franceschi.
\newblock Scaling of sub-gap excitations in a superconductor-semiconductor
  nanowire quantum dot.
\newblock {\em arXiv preprint arXiv:1609.07582}, 2016.

\bibitem{matsuura1977effects}
Tamifusa Matsuura.
\newblock The effects of impurities on superconductors with kondo effect.
\newblock {\em Progress of Theoretical Physics}, 57(6):1823--1835, 1977.

\end{thebibliography}
\newpage

\noindent \textbf{\large Supporting Information: Proximity-induced Shiba states in a molecular junction}

\setcounter{equation}{0}
\setcounter{figure}{0}
\setcounter{table}{0}
\setcounter{page}{1}
\setcounter{section}{0}

\section{Radical Polychlorotriphenylmethyl (PTM) Molecule}
As a prototypical spin 1/2 impurity for this study, we have judiciously selected a neutral and stable all organic radical (polychlorotriphenylmethyl, PTM) which we have shown produces a robust Kondo effect in gold break-junctions\cite{frisenda2015kondo}. The chemical structure of the molecule is shown on the left side of fig. 1(a) of the main text. Three chlorinated phenyl rings are connected to the central carbon atom through sp$^2$ hybridization\cite{ballester1985inert}. A single unpaired electron, giving the molecule an intrinsic paramagnetic ground state, is mainly localized at the central carbon atom and protected from the environment by the more bulky chlorine atoms. This protection gives the radical its high chemical and thermal stability. Additionally, this stability has been shown to result in Kondo correlations at low temperatures in gold break junctions which are stable against mechanical and electrostatic variations\cite{frisenda2015kondo}. Density functional theory (DFT) calculations show that the Fermi energy (with gold electrodes) of the molecule sits within a sizable energy gap (2 eV) between the singly occupied molecular orbital (SOMO) and the lowest unoccupied molecular orbital (LUMO) creating a stable electronic configuration with a large ``charging energy'' in the ground state. This is further verified by the absence of resonant transport at low bias in the whole range of gate voltages that can be applied in our electromigrated break junctions\cite{frisenda2015kondo}. This makes the PTM radical an ideal molecule to explore the large $U$ limit of the superconductor-QD phase diagram which favors Shiba excited states for both the singlet and doublet ground states. 

\section{Low temperature characteristics of empty gaps}
Empty nanogaps were measured at low temperatures in order to characterize the superconducting proximity effect in the gold electrodes without molecules. In fig. S1(a) we show the low temperature (100 mK) differential conductance ($dI/dV_b$) as a function of voltage bias ($V_b$) for an empty junction (no molecule). The two conductance peaks at voltages of $\approx\pm 0.8$ mV signal the overlap of the induced quasiparticle peaks in the two gold leads (see schematic in fig. S1(d)) and gives us a rough estimate of the induced mini-gap. A mini-gap ($\delta$, see fig. 1(a) of the main text) exists in the proximity-induced superconductor with a magnitude related to the Thouless energy, $E_{th}=\hbar D/L^2$, where $D=0.026$ m$^2$/s is the diffusion constant for gold\cite{wolz2011evidence} and $L$ is the electrode length\cite{belzig1996local}. If we assume a symmetric left and right gold electrode with a tunnel barrier in between, the length of each electrode is $L\approx140$ nm. This gives an estimate of $\delta=0.87$ meV. This is roughly twice as large as the induced gap we estimate from the position of the quasiparticle peaks in fig. S1(a). The geometry of the electrodes are most likely responsible for this discrepancy. In fig. S1(c) we show an SEM image of the junction after measurement. The asymmetry in the length of the left and right leads as well as the reduced, point-like shape of the left electrode lead to reduced proximity induced gaps. The position of the nanogap depends on the electromigration process and is not always symmetrically created at the center of the nanoribbon which leads to asymmetric proximity induced density of states (DOS).  Nonetheless, estimates of the average induced gap ($\delta_{avg}$) for the 7 molecular junctions are in the range of 0.7 meV to 1.1 meV which agrees well with estimates of the Thouless energy above.  

As opposed to coupling directly to the bulk and having a BCS like DOS, our gold leads offer external tuning of the induced gap through application of a magnetic field. This allows \textit{in-situ} tuning of $\delta$ relative to the Kondo energy, $T_K$, and exploration of the phase transition between the singlet and doublet ground states in the molecular junctions. Fig. S1(b) shows the tuning of the proximity induced gap as a function of magnetic field. The critical field of the junction reaches 200 mT.  

\section{The normal state characterization}
In this section we show the normal state characterization of a representative molecular junction. By applying a small finite perpendicular magnetic field (200 mT, directed out of the page in fig. 1(b)) to the junctions, the proximity effect in the normal leads can be completely suppressed. The curve in the inset of fig. S2(a) shows the differential conductance ($dI/dV_b$) as a function of bias voltage in the normal state for device B. A zero-bias Kondo peak is observed which closely resembles measurements of the same molecule in all-gold junctions\cite{frisenda2015kondo}. This peak is characterized as a function of temperatures up to 4 K (fig. S2(a)) and magnetic fields up to 8 T (fig. S2(b)). We observe the typical exponential decrease of the Kondo peak height with increasing temperature and its splitting in high magnetic fields. From the temperature dependence of the peak height we estimate a Kondo temperature of 4.4 K and from the splitting we estimate a g-factor of 2.4, all consistent with Kondo correlations observed in all-gold junctions. As reported previously\cite{frisenda2015kondo}, we also observe little modulation of the peak in the normal state with electrostatic gating due to the sizable SOMO-LUMO gap (see below, section Gate dependence). 

From the linear conductance we furthermore estimate the asymmetry in the coupling ($\Gamma_L, \Gamma_R$) to the leads using\cite{parks2007tuning}: 
\begin{equation} \label{asym}
G = \frac{2e^2}{h}\frac{4\Gamma_L\Gamma_R}{(\Gamma_L + \Gamma_R)^2}f(T/T_K) + G_b, 
\end{equation}
where $f(T/T_K)=[1+T^2/T_K^2(2^{1/s}-1)]^{-s}$ with $s = 0.22$ the expected value from numerical renormalization group theory for spin 1/2.\cite{costi1994transport} From Equation \ref{asym} we find an asymmetry of $\Gamma_L/\Gamma_R\approx 5 \times 10^{-4}$. This asymmetry is predominant in electromigrated junctions due to the random orientation of the molecule in the nanogap resulting in strongly different coupling strengths to the left and right electrodes.

\section{Gate dependence}
We observe little modulation of the conductance with back gate voltage due to the sizable (2 eV) SOMO-LUMO gap of the radical molecule. Fig. S3(a) shows $dI/dV$ as a function of $V_b$ and $V_g$ for device E. With our accessible gate range, no degeneracy point is reached that would signal a shift to another charge state. The same is true in the normal state (see Fig. S3(b)). In fact, very little modulation of the conductance is observed (see Fig. S3(c)) overall which supports the robust Kondo effect and stable spin 1/2 nature of the molecule. 

\section{Temperature dependence of the anomalous zero-bias peaks}
At finite temperatures (~1 K) a zero-bias peak emerges in the conductance curves which grows with temperature. Devices E and F show the presence of a zero-bias peak (see Device overview section). Zero bias peaks also emerge for devices A and D at higher temperatures (see Fig. S4). A plausible explanation for this peak is that a mini Kondo peak emerges in the gap at higher quasiparticle filling. Further supporting this claim is Fig. S4(c) which shows calculated curves of the modified Anderson impurity model (see Theoretical Model section below) as a function of increased broadening in the leads. Similar to an increase of quasiparticles at higher temperatures, increased broadening leads to a zero-bias peak.  

\section{Theoretical Model}
As described in the main text, our organic radical molecules are perfect model spin 1/2 impurities and as such can be described by an Anderson Hamiltonian of the form

\begin{eqnarray}
\label{su4vs2::eq:HD}
H_{mol} = \sum_{\sigma=\uparrow,\downarrow}
\varepsilon_{\sigma} d_{\sigma}^\dag d_{\sigma}
+ Un_{\uparrow}n_{\downarrow} \,,
\end{eqnarray}
where $\varepsilon_{\sigma}$ is the single-particle energy level of the
localized state with spin $\sigma$, $d_{\sigma}^\dag$
($d_{\sigma}$) the fermion creation (annihilation) operator of the
state,
\begin{math}
n_{\sigma} = d_{\sigma}^\dag d_{\sigma}
\end{math}
the occupation, and $U$ the on-site Coulomb
interaction, which defines the
charging energy.
Due to the sizeable SOMO-LUMO gap of ~2 eV of our molecular system, we focus on the regime where the charging energy is much bigger than other energy scales. In this regime the Hamiltonian in Eq.~(\ref{su4vs2::eq:HD}) suffices to describe all relevant physics.

In the normal state (section III above), Kondo physics arises as a result of the interplay between the strong correlation in the molecule and the coupling of the localized electrons with the electrons in the contacts. In our case, these are described as two leads ($\alpha=L$ and $R$) which can be represented by a Hamiltonian of the form:
\begin{equation}
\label{su4vs2::eq:HC1}
H_\alpha =
\sum_{k_\alpha}\sum_\sigma\varepsilon_{k_\alpha,\sigma}
a_{k_\alpha\sigma}^\dag a_{k_\alpha\sigma}
\end{equation}

Tunneling between the molecule and the contacts is described by the Hamiltonian
\begin{equation}
\label{su4vs2::eq:HT1}
H_T =
\sum_{k_\alpha\sigma}
\left(V_{k_\alpha\sigma}
  a_{k_\alpha\sigma}^\dag d_{\sigma}
  + h.c.\right) 
\end{equation}

The total Hamiltonian is then given by
\begin{math}
H = H_L + H_R + H_T + H_{mol} 
\end{math} \,. For simplicity, we ignore the $k$- and $\sigma$-dependence of the tunneling amplitudes.
Therefore, we consider a simplified model with $V_{k_\alpha
\sigma}=V_{\alpha}/\sqrt{2}$ which defines the widths $ \Gamma_\alpha^N = \pi\rho_0 |V_\alpha|^2$, where $\rho_0 $ is the (normal) density of states in the reservoirs.

As discussed in the main text, bulk superconducting electrodes (gap $\Delta$) induce a proximity effect in the gold electrodes with mini-gap $\delta$. This proximity effect can be modeled as a modified  coupling to the reservoirs of the form
\begin{equation}
\Gamma_\alpha(E)=\Gamma_\alpha^NN_{s}(E),
\end{equation}
where $N_{s}(E)$ is the density of states of the proximitized contacts. As discussed in the main text, it is clear from the experimental data that a clean BCS density of states is not a good description of the experiments.
Instead, we use a Dynes expression of the form
\begin{equation} 
N_{s}(E,\gamma)=Re[\frac{|E|+i\gamma}{\sqrt{(|E|+i\gamma)^2-\delta^2}}],
\end{equation}
where $\gamma$ is a phenomenological broadening which takes into account a finite density of states inside the BCS gap.

We are interested in a limit where the Coulomb interaction is very large such that we can safely take the limit of $U\rightarrow \infty$. In this case, the problem can be solved within the so-called Non-Crossing approximation (NCA) \cite{langreth1991derivation,wingreen1994anderson,hettler1994nonlinear,hettler1998nonequilibrium,aguado2003kondo} generalized to the superconducting case \cite{clerk2000loss,sellier2005pi}. The NCA is the lowest order fully self-consistent non-pertubative diagrammatic expansion that one can write for the $U\rightarrow \infty$ limit of the problem. This approximation neglects vertex corrections in the diagrammatic expansions and, as a result, the NCA fails in describing the low-energy Fermi-liquid regime. Nevertheless, the NCA has proven to give reliable results even at temperatures well below the Kondo temperature (of the order of $T=10^{-2}T_K$) \cite{costi1996spectral}.

Without entering into much detail of the theory, we
just mention that we solve the problem by deriving self-consistent equations-of-motion
for the time-ordered double-time Green's function by using the Keldysh method \cite{devreeese2013linear}.  In the paper, we just show numerical results of the resulting coupled set of integral NCA equations for our problem and refer the interested reader to
Refs.~\cite{langreth1991derivation,wingreen1994anderson,hettler1994nonlinear,hettler1998nonequilibrium,aguado2003kondo} for details. In particular, the density of states is given by
\begin{equation}
\rho(\omega)=-\frac{1}{\pi}\sum_{\sigma}
\mathrm{Im}[A^{r}_{\sigma}(\omega)],
\end{equation}
where $A^{r}_{\sigma}(\varepsilon)$ is the Fourier transform of
the retarded Green's function. 
Following Meir and Wingreen in Ref. \cite{meir1992landauer}, the current is given by:
\begin{equation}
\label{current}
I_{\alpha\in \{ L,R \}}=-\frac{2e}{h}\sum_{\sigma}\int d\epsilon
\Gamma_\alpha(\epsilon)[2Im A^r_{\sigma}(\epsilon)f_\alpha(\epsilon)
+A^<_{\sigma}(\epsilon)].\nonumber\\
\end{equation}
with $A^<_{\sigma}(\epsilon)$ the Fourier transform of the lesser Keldysh Green's function and $f_\alpha(\epsilon)=\frac{1}{1+e^{\frac{(\epsilon-\mu_\alpha)}{kT}}}$ the Fermi-Dirac function at each reservoir held at a chemical potential $\mu_\alpha$ such that the applied bias voltage is defined as $eV=\mu_R-\mu_L$. Eq. (\ref{current}) neglects Andreev processes and is a good approximation for situations where Shiba states resulting from strong coupling between the molecule and one of the proximitized electrodes are weakly probed by the other proximitized electrode, as described in the main text.

In practice, we self-consistently solve the NCA integral equations until good numerical convergence is reached. All $dI/dV$ calculations presented in the main text are done for finite temperatures in the range $T\sim T_K$ and different values of $\delta$ and 
$\gamma$.

\section{Device overview}
Fig. S5 shows an overview of all 7 junctions measured showing symmetric conductance peaks in the superconducting state and a zero-bias peak in the normal state. From the top to the bottom, the devices are arranged according to their critical field $B_c$ of the Kondo peak except for device G which does not show a splitting because the critical field is larger than our experimental range. All devices were measured at $\approx 100$ mK except for devices E and F which were measured at 1.8 K. The left side panels in fig. S5 shows the differential conductance $dI/dV_b$ as a function of voltage bias $V_b$ for each device in the superconducting state (0 T, black curve) and the normal state (200 mT, red curve). The right side panels show $dI/dV_b$ as a function of $V_b$ and magnetic field.

\section{Quantum phase transition between the doublet and singlet ground states}
From the 7 devices studied in detail, we make here an estimate of the ratio of $k_BT_K/\delta$ at which the quantum phase transition occurs between the two ground states. While we do not have a direct measurement of the induced superconducting gap of the probe electrode ($\delta_L$ in the main text), we can still make a rough estimate of the bound state energy ($E_b$) from the average induced superconducting gap ($\delta_{avg}$). For consistency, the Kondo energies were all estimated from the critical field following the analysis in the main text for device A. In Fig. S6 we plot $E_b/\delta_{avg}$ versus $k_BT_K/\delta_{avg}$ for all devices. Four devices were found to have induced superconducting gaps which were greater than their respective Kondo energies and for these devices, following convention, we have assigned a positive bound state energy. Three devices were found to have induced gaps with energies less than their Kondo energies. These devices we have assigned a negative bound state energy. Between the positive and negative bound state energies, a quantum phase transition occurs which is labeled in Fig. S6. We estimate the transition occurs for $0.9<k_BT_K/\delta_{avg}<1.1$. 

\clearpage

\begin{figure}
\centerline{\includegraphics[width=5in]{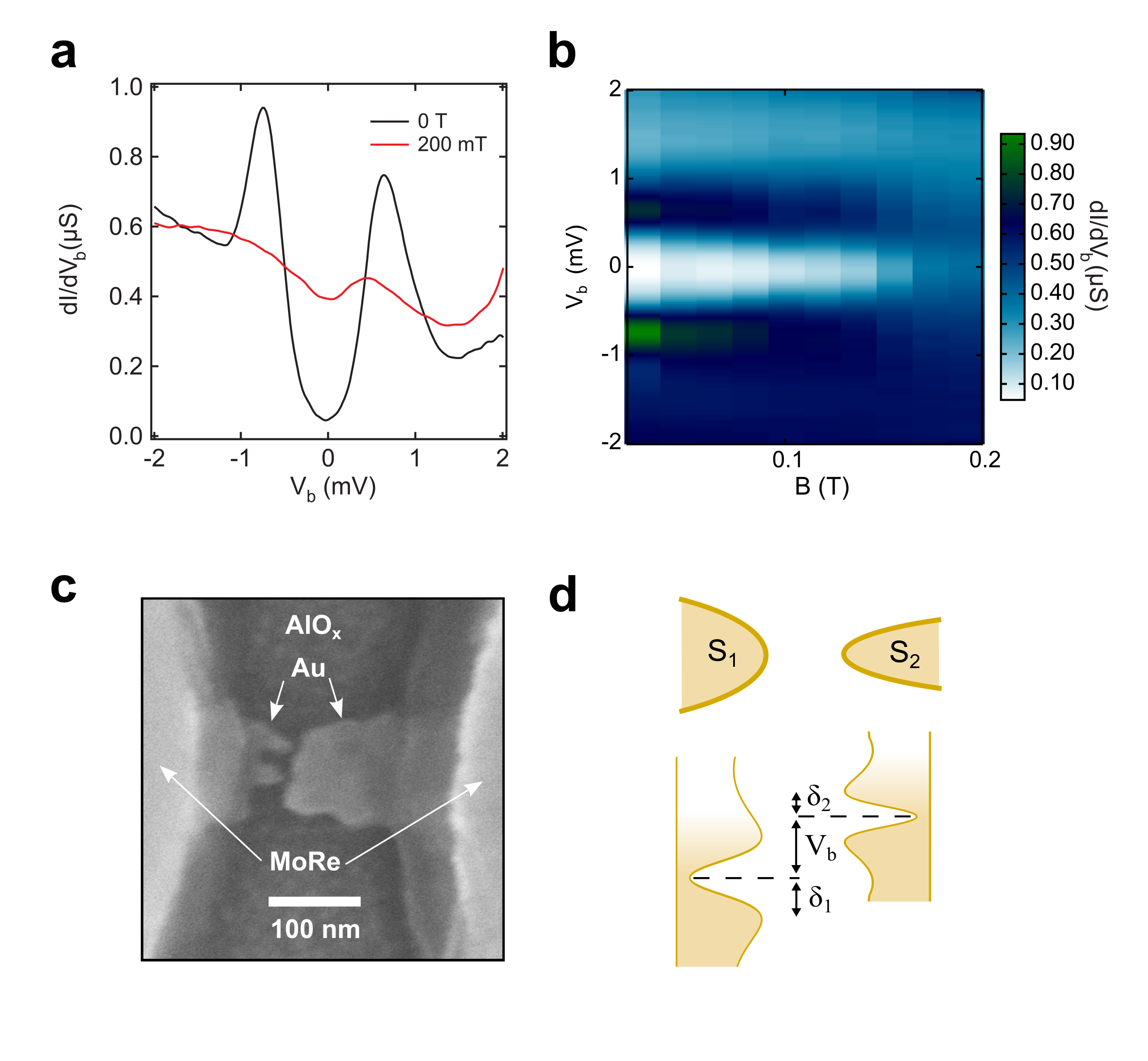}}
\justify FIG. S1. \textbf{Characteristics of an empty gap measured at low-temperature.} (a) Low temperature (100 mK) measurement of the differential conductance ($dI/dV$) as a function of voltage bias ($V_b$) of an empty gap. (b) Schematic representation of the asymmetry in the proximity induced gaps of the left and right leads.  (a) Numerical differential conductance ($dI/dV_b$) as a function of voltage bias ($V_b$). The black curve shows the measurement at zero field and the red curve shows the measurement near the critical field of the proximity induced superconductivity. (b) Color plot of $dI/dV_b$ as a function of $V_b$ and perpendicular magnetic field. (c) Scanning electron microscopy image of the junction after measurement. (d) Schematic of the density of states (DOS) of the two gold leads.   
\end{figure}

\begin{figure}
\centerline{\includegraphics [width=4.5in]{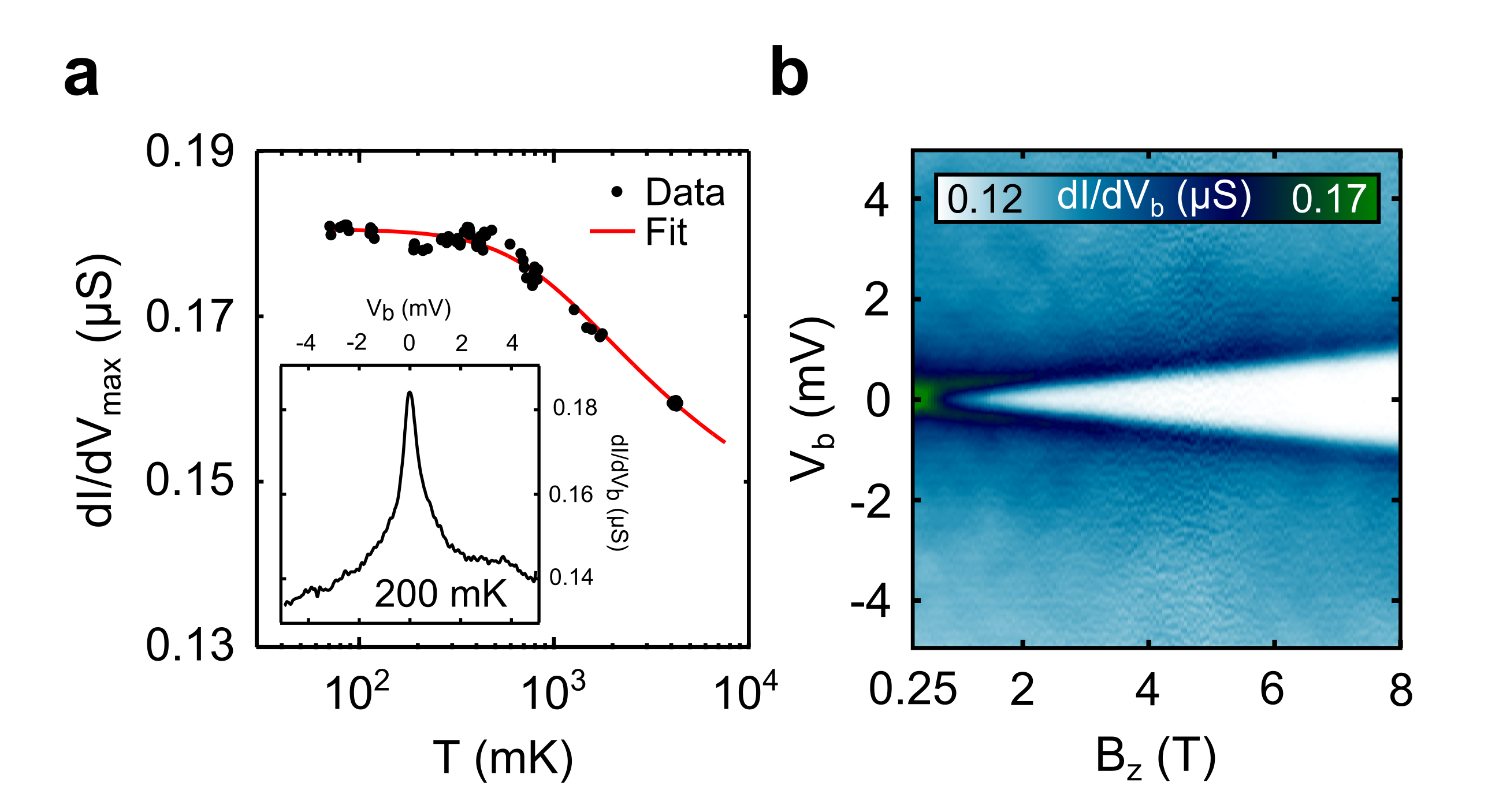}}
\justify FIG. S2. \textbf{Molecular junction in the normal state.} (a) Temperature dependence of the peak conductance which corresponds to a Kondo temperature of 4.4 K. The inset shows $dI/dV_b$ vs. $V_b$ showing the Kondo peak in the normal state (200 mT). (b) $dI/dV_b$ as a function of $V_b$ and magnetic field showing the splitting of the Kondo peak.
\end{figure}

\begin{figure}
\includegraphics[width=6in]{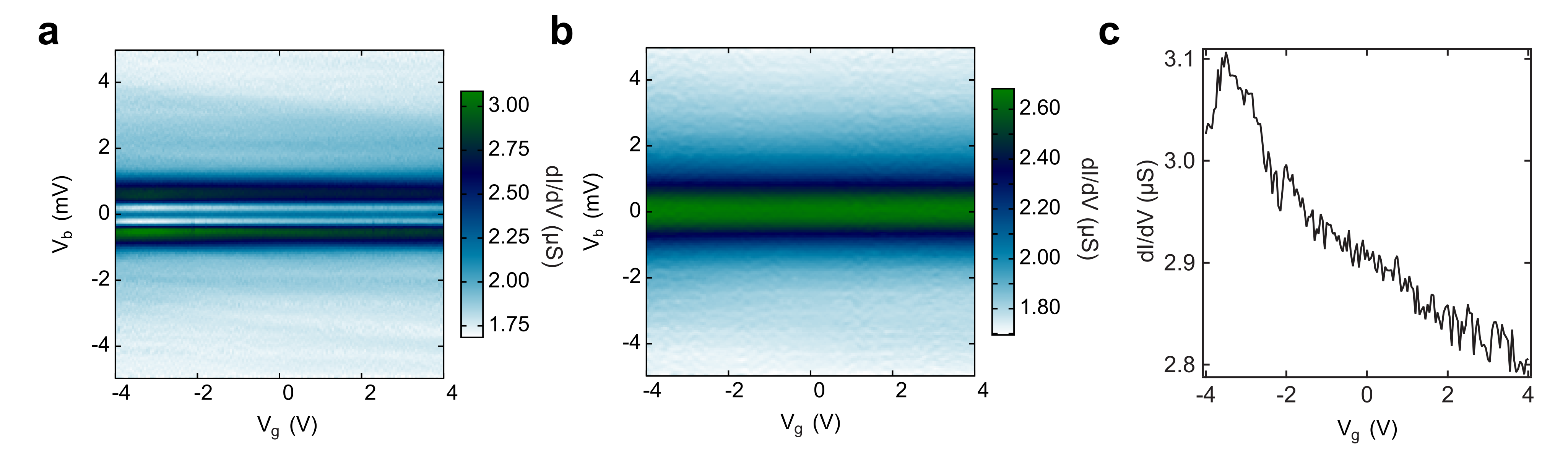}
\justify FIG. S3. \textbf{Gate dependence of a molecular junction.} (a) A color plot of the $dI/dV$ as a function of $V_b$ and $V_g$ in the superconducting state ($B = 0$ T). (b) A color plot of the $dI/dV$ as a function of $V_b$ and $V_g$ in the normal state ($B = 300$ mT). (c) Line cut from panel (a) at a voltage of $V_b = -0.5$ mV showing a small modulation with gate voltage. 
\end{figure}

\begin{figure}
\includegraphics[width=6in]{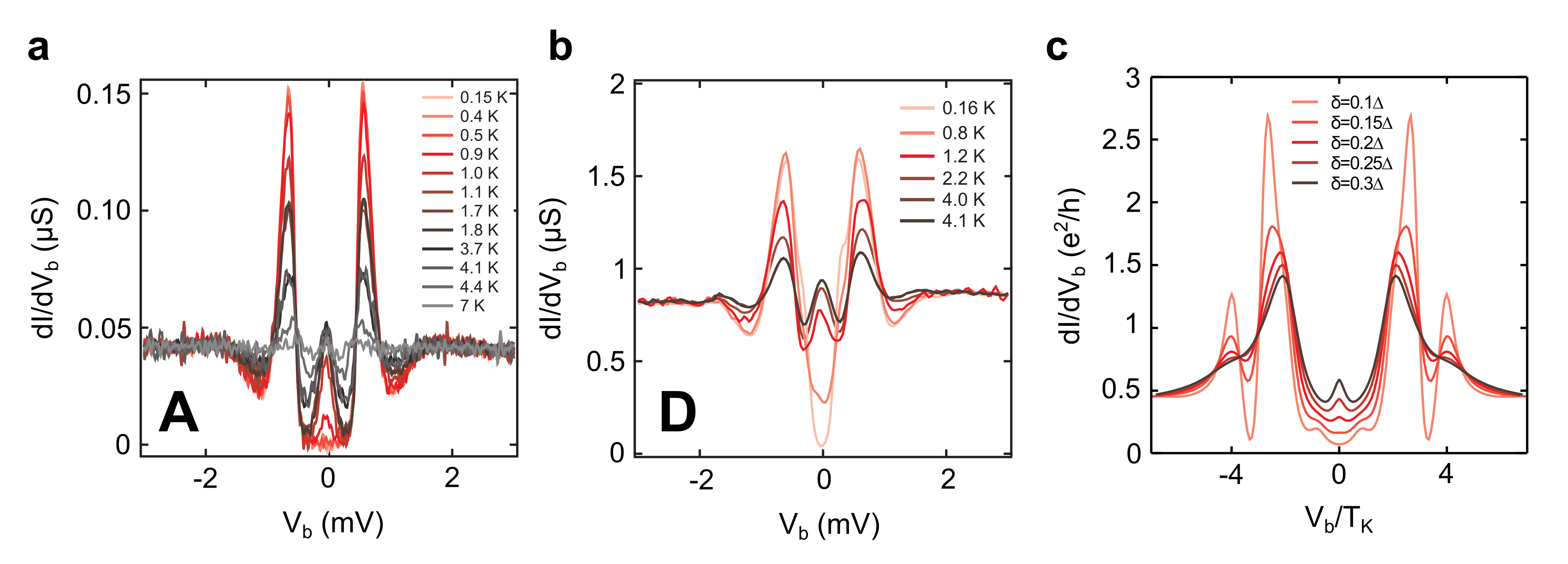}
\justify FIG. S4. \textbf{Temperature dependence of the anomalous zero-bias peak.} (a) $dI/dV$ as a function of $V_b$ for sample A as a function of temperature. (b) $dI/dV$ as a function of $V_b$ for sample D as a function of temperature. (c) Calculations of conductance as a function of $V_b$ for increasing phenomenological broadening ($\gamma$) of the BCS gap. 
\end{figure}

\begin{figure}
\centerline{\includegraphics[width=5in]{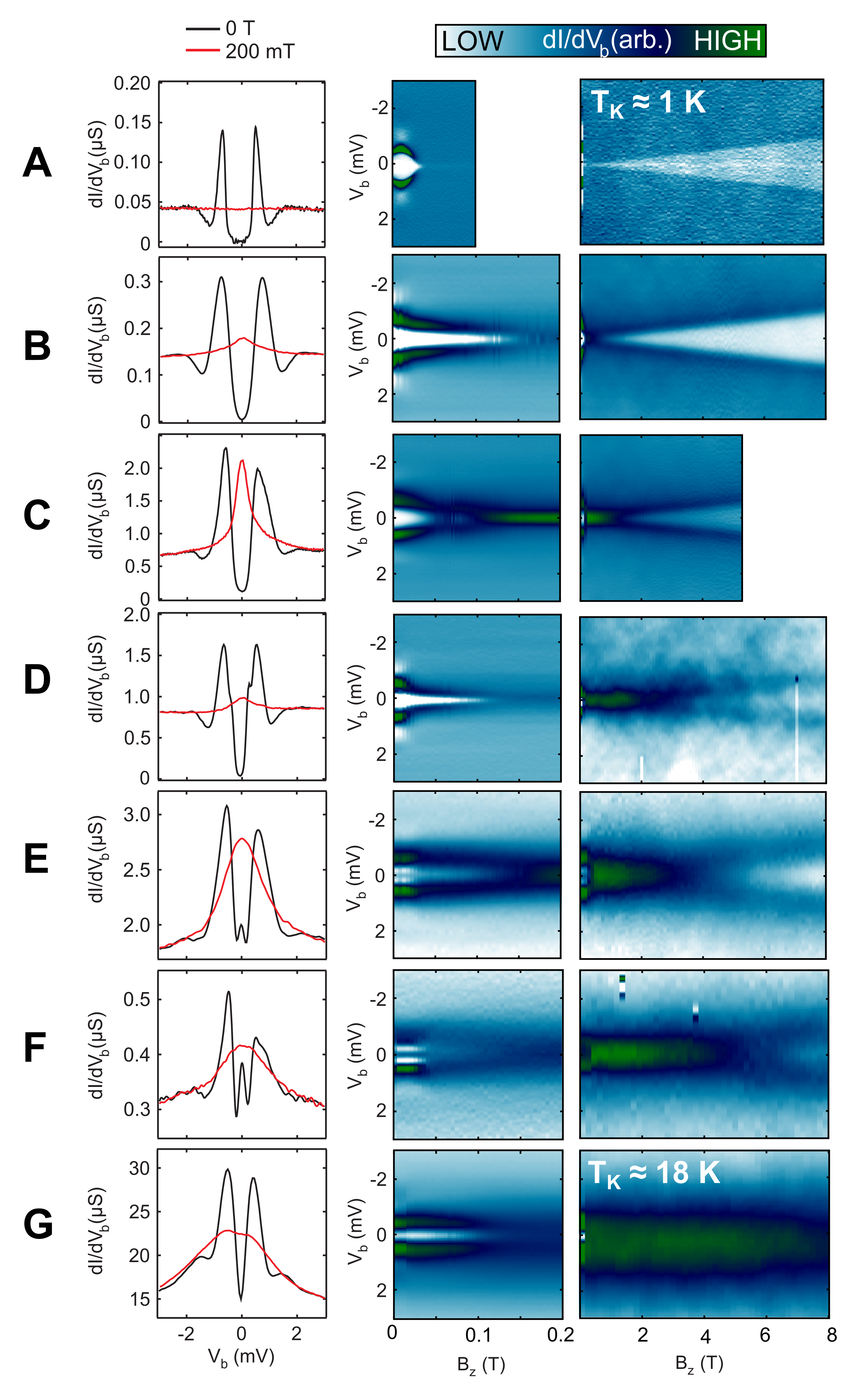}}
\justify FIG. S5. \textbf{Overview of all 7 devices measured.} The left-side panels show $dI/dV_b$ as a function of $V_b$ and the right-side panels show color plots of $dI/dV_b$ as a function of $V_b$ and magnetic field applied perpendicular to the junction.
\end{figure}

\begin{figure}
\includegraphics[width=3in]{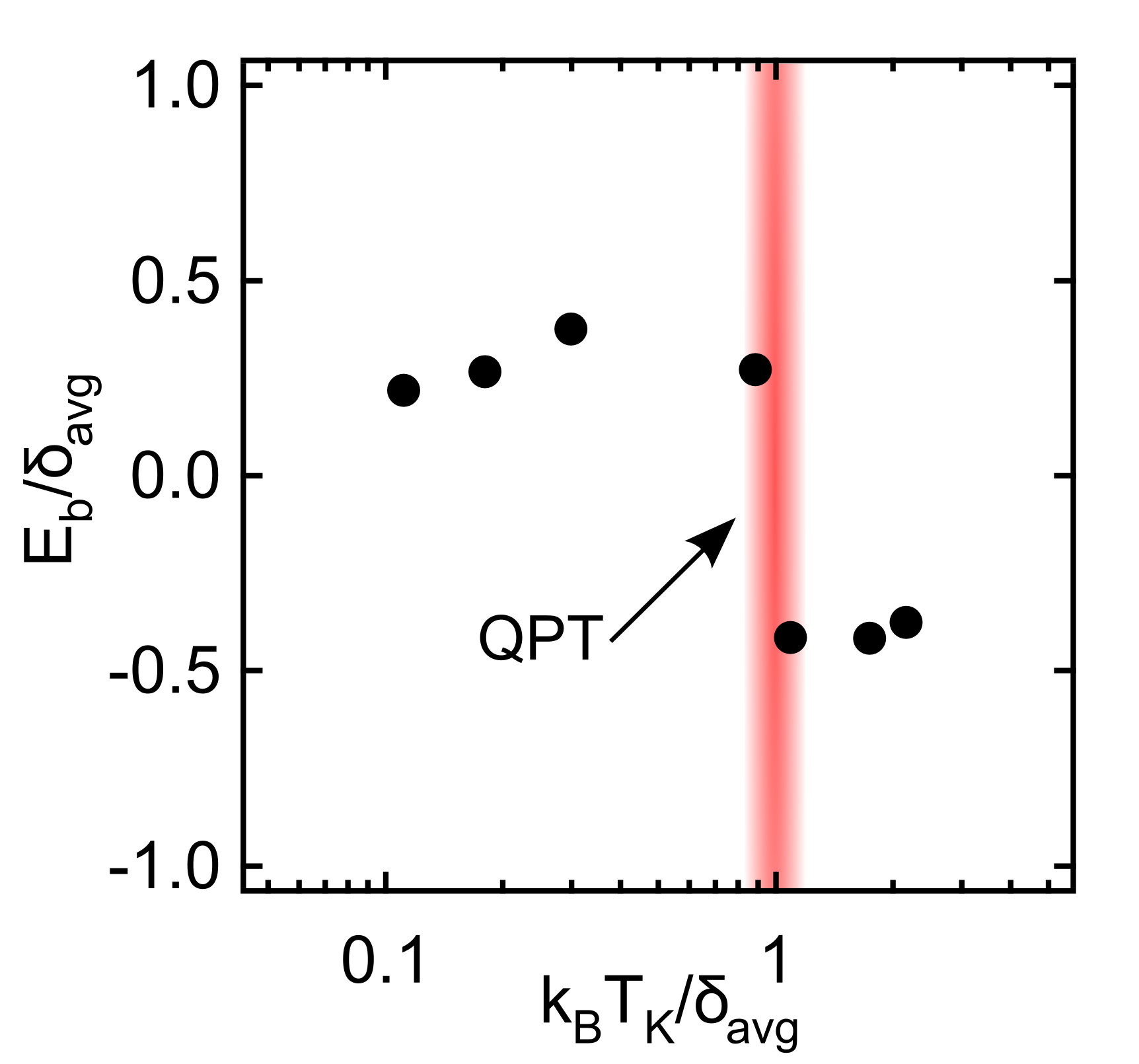}
\justify FIG. S6. \textbf{Quantum phase transition between the singlet and doublet ground states.} The bound state energies ($E_b$) for all seven devices plotted versus their respective Kondo energies ($k_BT_k$), both normalized by the average induced superconducting gaps ($\delta_{avg}$). 
\end{figure}
\clearpage

\end{document}